\documentclass[superscriptaddress,showpacs,amsmath,amssymb,twocolumn,nofootinbib]{revtex4-1}

\usepackage{amsmath, amsthm, amssymb, bm, braket}
\usepackage[dvips]{graphicx}
\usepackage{color}
\usepackage[normalem]{ulem}

\begin{document}
\title{Time-dependent Dirac equation applied to one-proton radioactive emission}

\author{Tomohiro Oishi}
\email[E-mail: ]{tomohiro.oishi@yukawa.kyoto-u.ac.jp}
\affiliation{Yukawa Institute for Theoretical Physics, Kyoto University, Kyoto 606-8502, Japan}

\renewcommand{\figurename}{FIG.}
\renewcommand{\tablename}{TABLE}

\newcommand{\red}[1]{\textcolor[rgb]{1.0,0.0,0.0}{\bf #1}}
\newcommand{\ora}[1]{\textcolor[rgb]{1.0,0.5,0.0}{\bf #1}}
\newcommand{\gre}[1]{\textcolor[rgb]{0.0,0.6,0.0}{\bf #1}}
\newcommand{\blu}[1]{\textcolor[rgb]{0.0,0.0,1.0}{\bf #1}}
\newcommand \modific[2]{#2} 

\newcommand{\crc}[1] {c^{\dagger}_{#1}}
\newcommand{\anc}[1] {c_{#1}}
\newcommand{\crb}[1] {a^{\dagger}_{#1}}
\newcommand{\anb}[1] {a_{#1}}

\newcommand{\CGC}[6] {\mathcal{C}^{(#1,#2){#3,#5}}_{#4,#6}} 
\newcommand{\JJJS}[6] {\left( \begin{array}{rrr} {#1}&{#3}&{#5}\\{#2}&{#4}&{#6} \end{array} \right)} 
\newcommand{\drsb}[2] {\bar{\psi}_{#1}(x_{#2})} 
\newcommand{\drsc}[2] {{\psi}^{\dagger}_{#1}(x_{#2})} 
\newcommand{\drsk}[2]     {{\psi}_{#1}(x_{#2})} 
\newcommand{\bi}[1] {\ensuremath{ \boldsymbol{#1} }}
\newcommand{\oprt}[1] {\ensuremath{ \hat{\mathcal{#1}} }}
\newcommand{\abs}[1] {\ensuremath{ \left| #1 \right| }}
\newcommand{\slashed}[1] {\not\!{#1}} 

\def \beq{\begin{equation}}
\def \eeq{\end{equation}}
\def \beqa{\begin{eqnarray}}
\def \eeqa{\end{eqnarray}}
\def \bir{\bi{r}}
\def \ubir{\bar{\bi{r}}}
\def \bip{\bi{p}}
\def \ubip{\bar{\bi{r}}}
\def \adel{\tilde{l}} 



\begin{abstract} \noindent
Relativistic energy-density functional (REDF) theory has been developed and utilized for self-consistent meanfield calculations of atomic nuclei.
The proton-emitting radioactivity can provide a suitable reference to improve the predicting ability of REDF especially on the proton-drip line.
One needs to consider the quantum tunneling effect, which plays an essential role in nucleon-emitting radioactive processes.
However, the relativistic quantum tunneling has been less investigated compared with the non-relativistic case.
This work is devoted to a theoretical evaluation of one-proton ($1p$) radioactivity based on the relativistic Dirac formalism.
For this purpose, I develop the time-dependent (TD) Dirac-spinor calculation to simulate the $1p$ emission.
By utilizing the relativistic Hartree-Bogoliubov (RHB) calculation with the DD-PCX parameters, single-proton potentials for the time-dependent Dirac spinor are determined.
The TD-Dirac calculation is applied to the $1p$ emissions from the $^{37}$Sc and $^{39}$Sc nuclei, which can be well approximated as the valence proton and the proton-close-shell cores.
The sensitivity of $1p$-emission energy and decaying width to the mass number is demonstrated.
Remarkable sensitivity exists due to the size of system, which affects the nuclear part of potentials and energy levels,
whereas the Coulomb barrier is common with the same atomic number.
The calculated $1p$ energy and decaying lifetime are roughly consistent to the experimental limitation.
The present TD-Dirac calculation is expected as applicable widely to proton-rich nuclides in order to improve the REDF by utilizing the $1p$-emission data.
\end{abstract}

\maketitle

\section{Introduction} \label{Sec:Intro}
Relativistic energy-density functional (REDF) theory has been one of the most successful frameworks to describe the physical properties of atomic nuclei \cite{1999Kohn_rev,1974Walecka,1984Serot,1989Reinhard,1996Ring_rev}.
For static properties, e.g. the binding energy and density distribution in the ground state of nucleus,
the self-consistent meanfield calculation based on the REDF theory has been utilized with fruitful results \cite{2005Vret,2006Meng,2011Niksic_rev}.
On the other hand, the dynamic properties have been less investigated.
The one-proton ($1p$) radioactivity belongs to this category.
In the static REDF-meanfield framework, nucleons are described with the Dirac equation self-consistently to their density distributions.
Even though there are various REDFs in the market,
for the proton-emitting radioactivity along the proton-drip line,
their ability and accuracy have not been sufficiently examined \cite{1998Vret_1p,2001Lala_1p}.
There has been a problem of the evaluation of decaying width or equivalently lifetime,
to which the pure-static calculation is not applicable.

For the description of proton emission with quantum-tunneling effect,
one needs some additional protocol \cite{47Krylov,89Kuku,72Tayler}, e.g. the time-dependent (TD), scattering theory, or non-Hermitian method.
This work employs the first option, which is suitable to simulate the dynamics.
That is, within the TD calculation, the behaviour of emitted particle(s) can be intuitively understood by following their time evolutions.
The decaying width or equivalently lifetime is directly evaluated from the time-dependent tunneling process.
The TD calculation has been utilized to describe a variety of nuclear meta-stable states \cite{87Gur,88Gur,94Serot,94Car,98Talou,00Talou,04Gur,2012Maru}, but mostly in the non-relativistic Schr\"{o}dinger formalism.
The relativistic version of $1p$-emission calculation involving the quantum-tunneling effect
has been on demand for the improvement of nuclear REDF theory by utilizing the $1p$-emission data.

As complementary options to the TD calculation, in nuclear physics, there have been other methods, namely,
the non-Hermitian \cite{2014Myo_rev,2022Myo_rev,2016Niu_Dirac,2017Fang_Dirac,2019Wang_Dirac,2022Cao}
and scattering-theoretical calculations \cite{72Tayler,1947Wigner,1987Hale,1988Shepard,2009Daoutidis,2010Li_Dirac,2014TTSun,2016TTSun,2020TTSun}.
Several non-Hermitian calculations have been utilized in
the non-relativistic \cite{1968Berggren,2014Myo_rev,2022Myo_rev,06Hagen,2002Michel,2017Tian} and
relativistic cases \cite{1968Berggren,2016Niu_Dirac,2017Fang_Dirac,2019Wang_Dirac,2022Cao} in order to describe nuclear meta-stable states.
Especially in Refs. \cite{2016Niu_Dirac,2017Fang_Dirac,2019Wang_Dirac,2022Cao},
the complex-momentum representation combined with the REDF calculation has been utilized.
In Ref. \cite{2016Niu_Dirac}, it is shown that the Dirac equation in the complex-momentum representation
enables one to solve the nuclear bound and resonant states on equal footing.
In Ref. \cite{2017Fang_Dirac}, the sensitivity of resonant energies and widths to the deformation
of the $^{37}$Mg nucleus is also discussed.
In Refs. \cite{1988Shepard,2009Daoutidis,2010Li_Dirac,2014TTSun,2016TTSun,2020TTSun},
the scattering-theoretical method within the Dirac-spinor formalism and/or REDF framework with continuum has been developed.
Recently, the evaluation of nucleon's resonance with the REDF and Green's function method has been performed in the neutron-rich side \cite{2020TTSun} and the proton-rich side \cite{2016TTSun}.
In Ref. \cite{2016TTSun}, it is concluded that the radius of relativistic meanfield potential plays the most important role to determine the proton's resonance.
In addition, for solving general quantum-resonant states,
the stabilization method with graph fitting has been also utilized \cite{1970Hazi,2011Ghoshal}.
Having a variety of options introduced above, however, there are still few REDF studies in the proton-rich side \cite{1998Vret_1p,2001Lala_1p,05Paar}.

\modific{1a}{For atomic nuclei, both the non-relativistic \cite{72Vautherin,03Bender_rev,UNEDF,1980Gogny,2009Goriely} and relativistic \cite{1974Walecka,1984Serot,1989Reinhard,1996Ring_rev,2005Vret,2006Meng,2011Niksic_rev} meanfield frameworks have been utilized.
In the non-relativistic (relativistic) side, nucleons are described with the Schr\"{o}dinger (Dirac) equation.
It is worth mentioning that the spin-orbit (LS) splitting is automatically concluded from the Dirac equation \cite{2000Greiner},
whereas the Schr\"{o}dinger equation includes extra LS parameter(s).
In principle, one can (cannot) separately deal with the LS and other parameters in the non-relativistic (relativistic) cases.
In the REDF-meanfield calculations, the spin degrees of freedom are naturally involved with the Dirac spinor, and the unified treatment of time-even and time-odd components of energy-density functionals can be guaranteed \cite{2005Vret,2006Meng}.
Even though these differences exist,
for reproducing the stable nuclei, the two frameworks reach the consensus in many cases \cite{2014Niksic,HFODD,HFBTHO2,2007Hila}.
For the proton-drip line, in contrast, finite ambiguities still remain depending on the choice of frameworks and parameters \cite{1998Vret_1p,2001Lala_1p,1984Jacek,96Jacek,13Olsen}.
}

In this work, I implement the TD calculation based on the Dirac-spinor formalism to describe the $1p$-emitting radioactivity.
The TD-Dirac calculation is then applied to the $1p$-emitting nuclei,
$^{37}$Sc and $^{39}$Sc, for benchmark of this method.
The sensitivity of $1p$-tunneling effect as well as decaying width to the mass number is also discussed.

In Sec. \ref{Sec:Form}, the basic formalism is introduced.
Sec. \ref{Sec:Results} is devoted to the numerical setting, results, and physical discussions.
Finally in Sec. \ref{Sec:Sum} I summarize this work.
In Appendix \ref{sec:complexscale}, the complex-scaled Dirac equation is utilized for the complementary calculation to the TD-Dirac one.
\modific{3}{In Appendix \ref{sec:initialtest}, the sensitivity of TD-Dirac results to the initial condiction is examined.}
I employ the CGS-Gauss system of units.
The spherical symmetry is assumed.

\section{Formalism} \label{Sec:Form}
\subsection{Dirac equation for spherical systems}
In this work, I focus on the quantum-tunneling process described by the Dirac equation \cite{2000Greiner}.
The single-particle (SP) Dirac equation for the valence proton $\psi(t,\bir)$ is given as
\beqa
 i\hbar c \frac{\partial}{\partial (ct)} \psi(t,\bir)
 &=& \Bigl[ -i\hbar c \beta \vec{\gamma} \cdot \vec{\nabla} +\beta Mc^2 \Bigr. \nonumber  \\
 && \Bigl.  +\beta S(r) +W(r) \Bigr] \psi(t,\bir), \label{eq:V7AF9}
\eeqa
where $M$ indicates the proton mass.
Here $S(r)$ and $W(r)$ are the scalar and vector potentials, respectively \cite{1974Walecka,1984Serot,1989Reinhard,1996Ring_rev}.
Note that $W(r)$ also includes the Coulomb potential originating from the photon field.
For the static solution, which satisfies $i\hbar \partial_{t} \psi_{N} =E_N \psi_{N}$,
the Dirac equation is simplified as
$\oprt{H}_{D} \psi_{N}(t,\bir) = E_{N} \psi_{N} (t,\bir)$,
where $\oprt{H}_{D}$ is the Dirac Hamiltonian:
\beq
 \oprt{H}_{D} \equiv -i\hbar c \beta \vec{\gamma} \cdot \vec{\nabla} +\beta Mc^2 +\beta S(r) +W(r). \label{eq:VT26WP}
\eeq
In this paper, the problem is limited to spherical systems.
The spherical Dirac spinor has the quantum labels of $N=\left\{ nljm \right\}$, including
the node number $n$, orbital angular momentum $l$, coupled angular momentum $j$,
and magnetic quantum number $m$.
This SP spinor $\psi_{N}$ is generally formulated as \cite{2000Greiner}
\beq
 \psi_N(\bir)
 = \left( \begin{array}{r}  iF_N(\bir) \\~~~\\ G_N(\bir) \end{array} \right) 
 = \left( \begin{array}{r}  i \frac{a_{nlj}(r)}{r} {\mathcal Y}_{ljm}(\ubir) \\~~~\\ \frac{b_{nlj}(r)}{r} \frac{\vec{\sigma}\cdot \bir}{r} {\mathcal Y}_{ljm}(\ubir) \end{array} \right), 
\eeq
where the angular part reads ${\mathcal Y}_{ljm}(\ubir) = \left[Y_{l}(\ubir) \otimes \chi \right]_{jm}$
with $\hat{s}_z \chi_{\pm 1/2}=\pm \frac{1}{2} \chi_{\pm 1/2}$ for the spin component.
Note that $\frac{\vec{\sigma}\cdot \bir}{r} {\mathcal Y}_{ljm}(\ubir) = {\mathcal Y}_{\adel jm}(\ubir)$,
where $\adel =l \mp 1$ when $l=j \pm \frac{1}{2}$.
By using this ansatz, the matrix equation for the larger component $a_{nlj}(r)$ and smaller component $b_{nlj}(r)$ can be obtained as
\beqa
 \left[ \frac{d}{dr}-\frac{\kappa_{lj}}{r} \right] a_{nlj}(r) &=& \frac{Mc^2+S(r)+E_N-W(r)}{\hbar c} b_{nlj}(r), \nonumber  \\
 \left[ \frac{d}{dr}+\frac{\kappa_{lj}}{r} \right] b_{nlj}(r)  &=&  \frac{Mc^2+S(r)-E_N+W(r)}{\hbar c} a_{nlj}(r), \nonumber
\eeqa
where $\kappa_{lj} = l+1$ for $j=l+1/2$ and $\kappa_{lj} = -l$ for $j=l-1/2$.
By introducing the new symbols as $s(r)\equiv Mc^2+S(r)$ and $v(r,E_N) \equiv E_N-W(r)$,
then the last equation for $\{a_{nlj}(r),b_{nlj}(r)\}$ can be simplified as
\beq
\frac{d}{dr}
\left( \begin{array}{c} a_N \\ b_N \end{array}  \right)
= \left( \begin{array}{cc} \frac{\kappa_{lj}}{r} & \frac{s+v}{\hbar c}  \\  \frac{s-v}{\hbar c} & \frac{-\kappa_{lj}}{r}  \end{array} \right)
\left( \begin{array}{c} a_N \\ b_N \end{array} \right). \label{eq:CF29KQ}
\eeq
This matrix equation is numerically solved with the Runge-Kutta method \cite{1989Atkinson} in this work.
The asymptotic form of $a_{nlj}(r)$ at $r\cong 0$ is given as
\beq
 a_{nlj}(r\cong 0) = r^{l+1} + \frac{C(r)}{4l+6}r^{l+3} + \mathcal{O}(r^{l+5}).
\eeq
where $C(r) \equiv \frac{s^2(r)-v^2(r)}{(\hbar c)^2}$.
The corresponding $b_{nlj}(r)$ can be computed as
\beq
 b_{nlj}(r\cong 0) = \frac{\hbar c}{s(r)+v(r)} \left[ \frac{da_{nlj}}{dr} -\frac{\kappa_{lj}}{r}a_{nlj}(r) \right].
\eeq
Note that I discuss only the case where the potentials vanish at $r \longrightarrow \infty$ in this paper.
In the non-relativistic limit, Eq. (\ref{eq:CF29KQ}) reduces to the
Schr\"{o}dinger equation including the potential term, $S(r)+W(r)$ \cite{1984Serot,2000Greiner}.

Equation (\ref{eq:CF29KQ}) needs the single-particle (SP)
potentials $S(r)$ and $W(r)$ as input. 
In this work, these potentials are determined by solving the 
self-consistent meanfield calculation, namely, the relativistic Hartree-Bogoliubov (RHB) calculation
for the system of interest \cite{2005Vret,2006Meng,2011Niksic_rev,2014Niksic}.
The setting of RHB calculation is presented in the next section.

\subsection{Time-dependent calculation}
For the nucleon-emitting process, I employ the TD calculation combined with the confining potential.
There have been several works, where a similar confining procedure but of the non-relativistic version is utilized to describe the quantum tunneling process as well as meta-stable state \cite{87Gur,88Gur,94Serot,94Car,98Talou,00Talou,04Gur,2012Maru}.
For fixing the initial state $\psi(t=0,\bir)$ of the Dirac tunneling state, 
a confining Hamiltonian is employed: $\oprt{H}'_{D} \equiv \oprt{H}_{D} + \beta \Delta S(r) + \Delta W(r)$,
where $\oprt{H}_{D}$ is the original Dirac Hamiltonian in Eq. (\ref{eq:VT26WP}).
The confining potentials, $\beta \Delta S(r) + \Delta W(r)$, will be determined so as to realize that the initial state can be well localized inside the potential barrier.
Its details are presented in the next section with numerical results.

When the initial state is determined, that can be expanded on the eigenstates of the original Hamiltonian. 
Namely, $\ket{\psi(t=0)} = \sum_{N} \alpha_{N} \ket{\psi_N}$,
where $\oprt{H}_{D} \ket{\psi_N} = E_N \ket{\psi_N}$.
Then the time evolution can be simply computed as
\beq
\ket{\psi(t)} = \exp \left[ -it\frac{\oprt{H}_{D}}{\hbar} \right] \ket{\psi(0)} = \sum_{N} e^{-it E_N/ \hbar} \alpha_{N} \ket{\psi_N}.
\eeq
Note that continuum states with $E_N >0$ are discretized within the finite box $R_{\rm max}$.
For a sufficiently large box, e.g. $R_{\rm max} \ge 150$ fm, I have checked that results in the following sections do not change, but except in the long-time region, where the contamination by reflected waves occurs.

\begin{table}[b]
\caption{Parameters used for Dirac SP potentials in Eqs. (\ref{eq:swform1})-(\ref{eq:swform3}).}\label{table:SWPC}
\begin{tabular*}{\hsize} { @{\extracolsep{\fill}} lclcll }
\hline  \hline
&&$^{36}$Ca$+p$  &&$^{38}$Ca$+p$ &unit \\
\hline
~$V_S,~V_W$ &&$-396.303,~355.082$   &&$-394.327,~356.659$ &[MeV]   \\
~$d_S,~d_W$  &&$3.72411,~3.68566$   &&$3.83319,~3.75805$  &[fm]  \\
~$a_S,~a_W$  &&$0.537403,~0.518423$   &&$0.52289,~0.512721$  &[fm]  \\
~$U_S,~U_W$  &&$-83.7739,~41.2407$   &&$-82.1177,~40.0851$ &[MeV]  \\
~$g_S,~g_W$  &&$0.145244,~0.374108$   &&$0.155998,~0.574995$ &[fm$^{-2}$]  \\
~$r_C$    &&$3.63212$   &&$3.69817$  &[fm]  \\
\hline  \hline
\end{tabular*}
\end{table}

\section{Results} \label{Sec:Results}
\subsection{Benchmark calculation for Sc-37}
In this section, I focus on the benchmark of one-proton ($1p$) emission from the $^{37}$Sc nucleus,
which is interpreted as the $^{36}$Ca$+p$ two-body system.
In order to determine the scalar and vector SP potentials, $S(r)$ and $W(r)$, I utilize the RHB calculation for this system \cite{2014Niksic}.
\modific{4a}{Because the $^{36}$Ca core is proton-shell closure at $Z=20$,
the RHB calculation reduces to the pure Hartree one in the proton side, i.e., the proton's pairing vanishes.
Also, the valence proton atop the Ca core is single without other protons in the same orbit.
Thus, I assume that the valence proton for emission can be well approximated by directly applying the RHB potential in this case.
This valence proton is expected to have a resonance in the $f_{7/2}$ channel.
}
In experimental data \cite{NNDC_Chart,2021Wang_AME}, the Q value of $1p$ emission is $2.9(3)$ MeV, whereas its lifetime or equivalently decaying width has not been measured.

In the present RHB calculation, I use the same setting as in Ref. \cite{2019Yuksel}.
Namely, the DD-PCX set of parameters is employed with the no-sea approximation.
Note that this setting for RHB has been a successful option to reproduce the ground-state properties of stable nuclei, including their binding energies, pairing gaps, and charge radii \cite{2019Yuksel,2021Perera}.

Since the scalar and vector potentials are solved as numerical data from RHB,
I employ the fitting functions to mimic them.
Their forms read
\beq
S(r)=\frac{V_S}{1+e^{\frac{r-d_S}{a_S}}} + U_S e^{-g_S r^2},\label{eq:swform1}
\eeq
as well as
\beq
W(r)=\frac{V_W}{1+e^{\frac{r-d_W}{a_W}}} + U_W e^{-g_W r^2} +V_C(r), \label{eq:swform2}
\eeq
where the Coulomb potential is also employed:
\beq
V_C(r)=\left\{ \begin{array}{ll}
    -\frac{Z e^2}{r} &~~~(r> r_C) \\
    -\frac{Z e^2}{2r_C} \left[ 3-\left( \frac{r}{r_C} \right)^2 \right] &~~~(r\le r_C) \end{array}
    \right. \label{eq:swform3}
\eeq
with $Z=20$.
Their parameters obtained by fitting are summarized in Table \ref{table:SWPC}.

\begin{figure}[tb] \begin{center}
\includegraphics[width = 0.99\hsize]{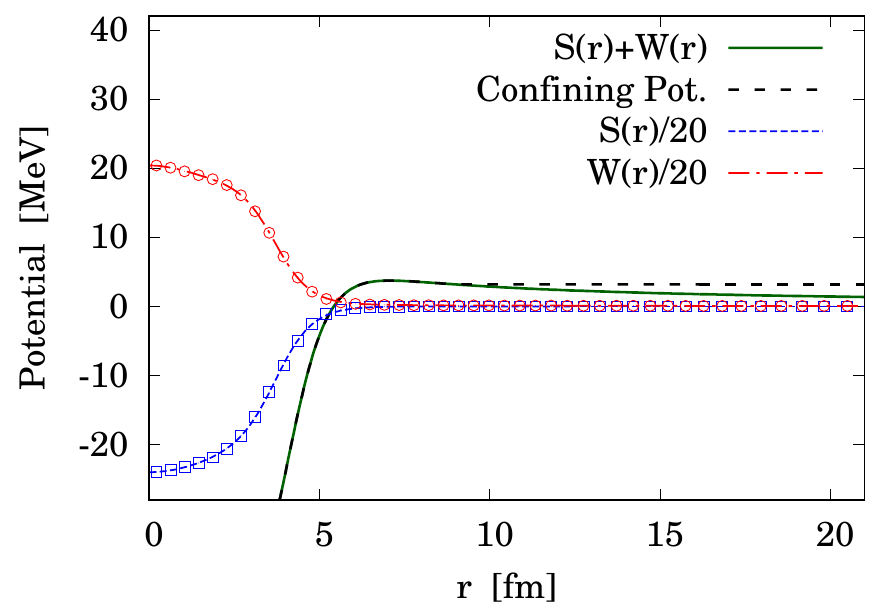}
\caption{The SP potentials $S(r)$ and $V(r)$ for the $^{37}$Sc $= ^{36}$Ca$+p$ system.
Those are fitted to mimic the RHB results of scalar and vector potentials, which are also plotted with square and circle symbols, respectively.
} \label{fig:6661}
\end{center} \end{figure}

The obtained potentials $S(r)$ and $W(r)$ from DD-PCX RHB are displayed in Fig. \ref{fig:6661} with the factor $1/20$ for plotting convenience. 
One can read that the total potential $S(r)+W(r)$ is determined as the small gap of two large quantities $S(r)$ and $W(r)$. 
There is also the barrier around $r\cong 6$ fm due to the Coulomb repulsive interaction between the valence proton and the $^{36}$Ca nucleus.
This barrier is essential for the $1p$ emission by the quantum-tunneling effect.

\begin{figure}[t] \begin{center}
  \includegraphics[width = 0.99\hsize]{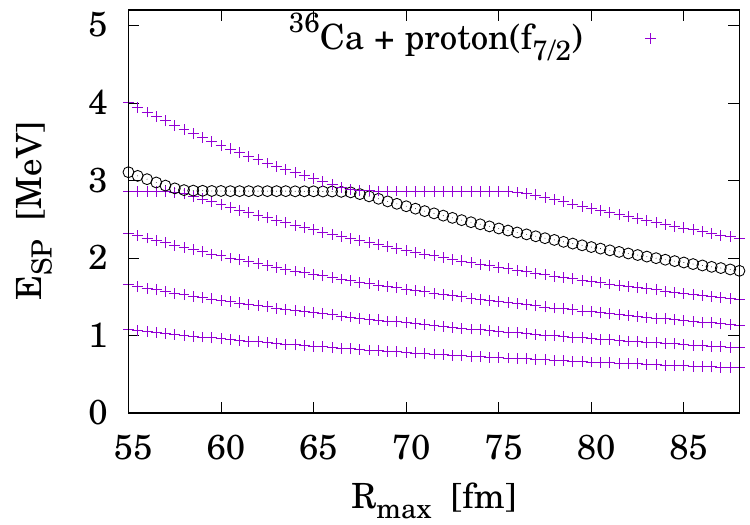}
  \includegraphics[width = 0.99\hsize]{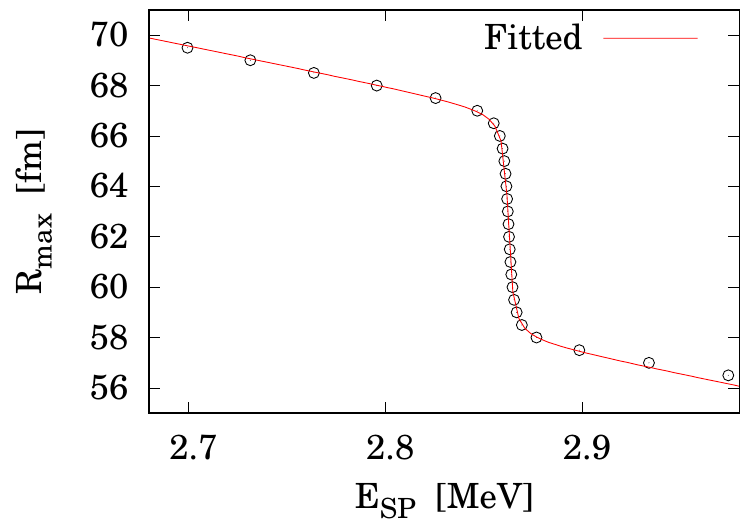}
  \caption{(Top) The continuum SP energies for the $^{37}$Sc $= ^{36}$Ca$+p$ calculation as the function of box size obtained with the $f_{7/2}$ ($l=3$ and $j=7/2$) setting.
The level used for fitting procedure is plotted with open circles.
(Bottom) The fitted function given in Eq. (\ref{eq:CUTEFS}).} \label{fig:7771}
\end{center} \end{figure}

Before the TD calculation, I check the possible resonant channels of the SP potentials in Fig. \ref{fig:6661}.
For this purpose, the stabilization technique is utilized \cite{1970Hazi,2011Ghoshal}.
That is, the SP energies in the continuum region ($E_{N} \ge 0$) are numerically solved by changing the radial-box size, $R_{\rm max}$.
Its result is presented in Fig. \ref{fig:7771}.
One reads that the proton state in the $f_{7/2}$ channel shows numerically the stable energy around $2.9$ MeV as the sign of resonance.
For the resonance pole of $\epsilon = Q_p -i\Gamma_p /2$,
with this stabilization graph, the resonance energy (real part) and width (imaginary part)
can be evaluated from the graphical fitting \cite{2011Ghoshal}.
That is, by using additional coefficients for smooth background,
\beq
\frac{dR_{\rm max}}{dE} \cong c_1 \frac{\Gamma_p /2}{(E-Q_p)^2 +(\Gamma_p /2)^2} +c_2,
\eeq
or equivalently,
\beq
R_{\rm max}(E) \cong  c_1 \arctan \left( \frac{E-Q_p}{\Gamma_p/2}  \right)  +c_2 E +R_0.\label{eq:CUTEFS}
\eeq
By fitting this Eq. (\ref{eq:CUTEFS}) to the result between $2.8-2.9$ MeV shown in Fig. \ref{fig:7771},
I obtained $Q_p =2.863$ MeV, $\Gamma_p =3.6181 \times 10^{-3}$ MeV, $c_1=-2.9146$ fm, $c_2=-15.77$ fm/MeV, and $R_0=107.6$ fm.
The fitted function is plotted in Fig. \ref{fig:7771}.
Note that this $Q_p$ value is consistent to the experimental data, namely $-S_{p}=2.9(3)$ MeV of $^{37}$Sc \cite{NNDC_Chart,2021Wang_AME}.

For the SP potentials in Fig. \ref{fig:6661},
I checked that the other channels do not show the stability, and thus, they are expected as non-resonant continuum.
I also confirmed that the bound states exist up to the $0d_{3/2}$ orbit for the $Z=20$ protons, and there are no other bound states.

\begin{figure}[t] \begin{center}
  \includegraphics[width = 0.99\hsize]{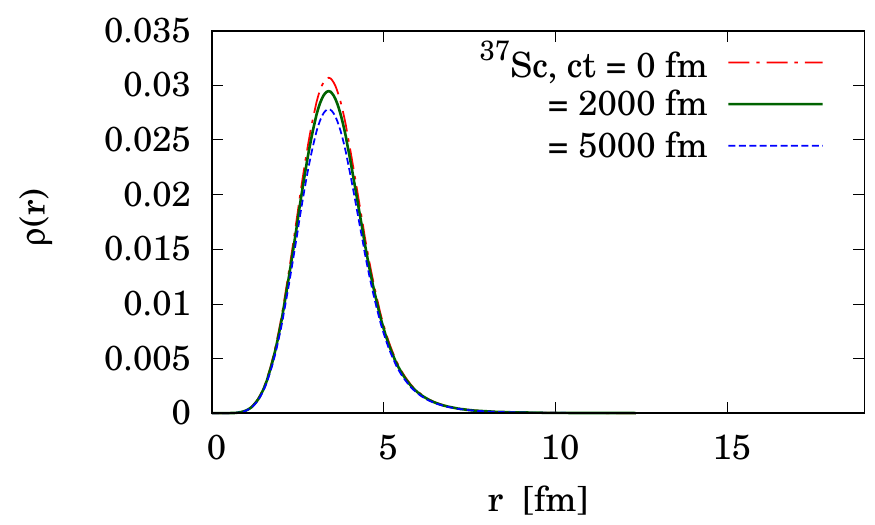}
  \includegraphics[width = 0.99\hsize]{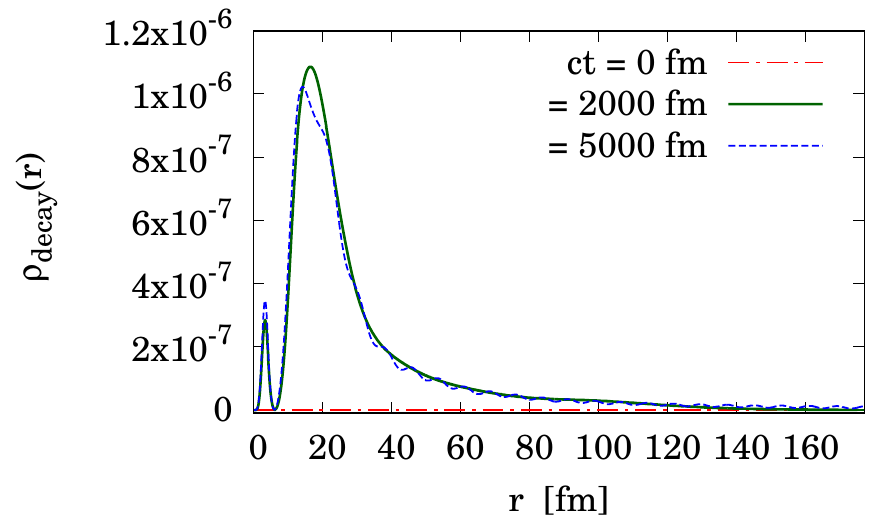}
  \caption{(Top) The one-proton density $\rho(t,r) = \bar{\psi}(t,r) \psi(t,r)$ during the time evolution in the $^{37}$Sc case.
(Bottom) Same plotting but for the decaying state $\rho_{\rm decay}(t,r) = \bar{\psi}_{\rm decay}(t,r) \psi_{\rm decay}(t,r)$.} \label{fig:0276}
\end{center} \end{figure}

\subsection{Time-dependent one-proton emission}
From this point I focus on the $f_{7/2}$ channel, which is expected as the resonance.
The confining potential to determine the initial state is plotted in Fig. \ref{fig:6661}.
Namely, I simply assume the wall potential for $r \ge 9$ fm.
The $1p$ energy of this initial state is calculated as $\Braket{\psi(0) \mid \oprt{H}_{D} \mid \psi(0)}=2.864$ MeV.
Thus, this setting is consistent to the experimental Q value of $^{37}$Sc, namely $-S_{p}=2.9(3)$ MeV \cite{NNDC_Chart,2021Wang_AME}.
In Fig. \ref{fig:0276}, the proton-density distribution of this initial state is displayed.
One can read that, at $t=0$, this initial state is well confined inside the potential barrier around $r\cong 6$ fm.
The box size is fixed as $R_{\rm max}=300$ fm.

For $t\ge 0$, as displayed in Fig. \ref{fig:0276}, the $1p$-density distribution decreases inside the barrier.
This behaviour is consistent to the quantum-tunneling picture.
For this tunneling process,
it is more convenient to focus on the decaying state.
That is,
\beq
 \ket{\psi_{\rm decay}(t)} = \ket{\psi(t)} -\beta(t) \ket{\psi(0)}, \label{eq:psidecay}
\eeq
where the survival coefficient, $\beta(t)$, is defined as the overlap between the initial and the present states.
That is, 
\beq
  \beta(t) \equiv  \Braket{\psi(0) | \psi(t)} = \sum_{N} \abs{\alpha_N}^2 e^{-itE_N/\hbar}. \label{eq:Krylov}
\eeq
Notice that $\Braket{ \psi (0) \mid \psi_{\rm decay}(t,r)}=0$, as well as $\psi_{\rm decay}(t=0,r)=0$ since $\beta(0) = 1$ from the initial normalization.
Namely, the decaying state $\psi_{\rm decay}(t,r)$ represents the deviation from the initial state.
In Fig. \ref{fig:0276}, the $1p$-density distribution of the decaying state is plotted.
The component outside the Coulomb barrier ($r \geq 6$ fm) remarkably increases along the time evolution,
that is consistent to the quantum-tunneling picture.

\begin{figure}[t] \begin{center}
  \includegraphics[width = 0.99\hsize]{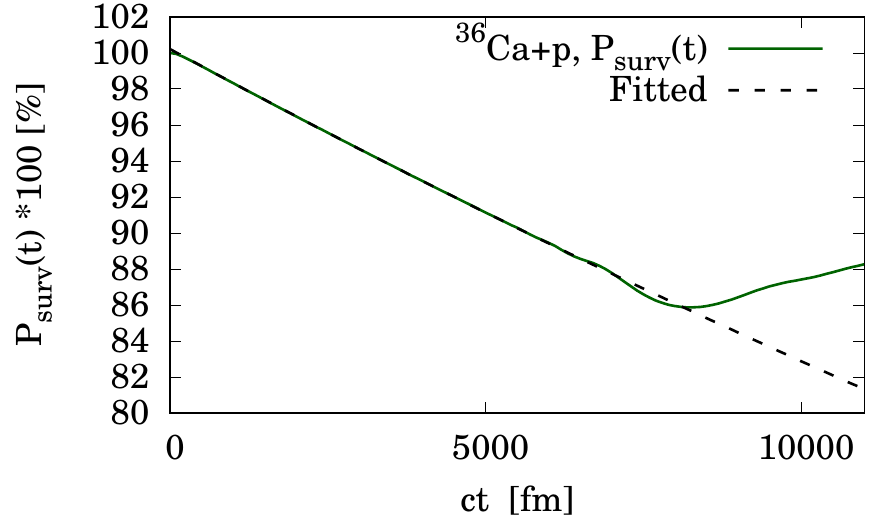}
  \includegraphics[width = 0.99\hsize]{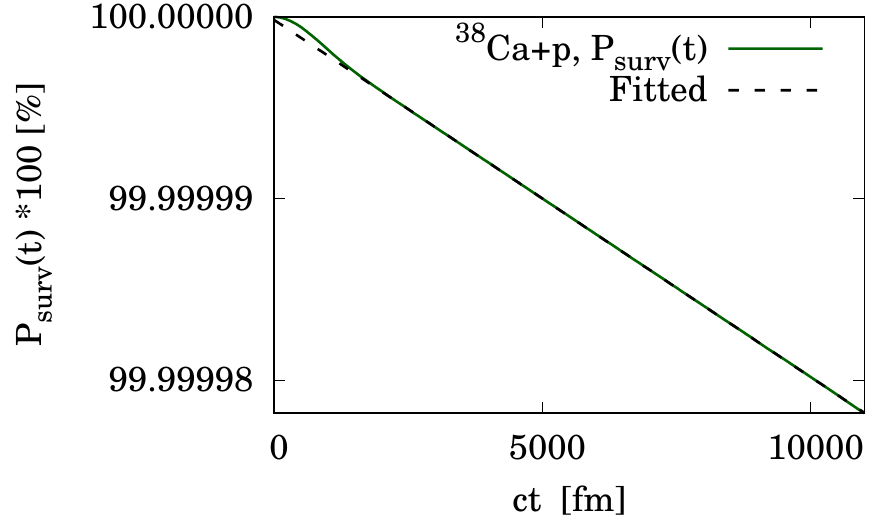}
  \caption{(Top) Survival probability $P_{\rm surv}(t)$ for $^{36}$Ca$+p$ in the $f_{7/2}$-resonant channel.
The fitted function is obtained as $f(t) =e^{-(ct) \Gamma_p / \hbar c} +0.00206$ with $\Gamma_p =3.74$ keV from the fitting procedure during $ct=1000-5000$ fm.
(Bottom) Same plotting but for $^{38}$Ca$+p$ in the $f_{7/2}$-resonant channel.
The fitted function is obtained as $f(t) =e^{-(ct) \Gamma_p/ \hbar c}$ with $\Gamma_p=3.87 \times 10^{-9}$ MeV.
} \label{fig:3702}
\end{center} \end{figure}

From Eq. (\ref{eq:Krylov}), one can read that 
the survival coefficient $\beta(t)$ is given by the Fourier transformation of the energy spectrum \cite{47Krylov,89Kuku}. 
The survival probability is then given as 
\beq
  P_{\rm surv}(t) = \abs{\beta(t)}^2, 
\eeq
which physically represents the radioactive-decaying rule from this initial state.
For example, when the state of interest has the Breit-Wigner (BW) spectrum with the width $\Gamma$,
its time evolution concludes the exponential-decaying rule: $P_{\rm surv}(t) \propto e^{-t/\tau}$ with the lifetime $\tau=\hbar /\Gamma$ \cite{47Krylov,89Kuku}.

\begin{table}[tb]
\caption{The Q value and decaying width of $1p$ emission evaluated in the $f_{7/2}$ channel of the $^{36}$Ca$+p$ case. Three methods utilized in this work are compared.
The unit is MeV.}\label{table:5324}
\begin{tabular*}{\hsize} { @{\extracolsep{\fill}} lccc}
\hline  \hline
~method &$Q_{p}$  &$\Gamma_{p}$  \\
\hline
~time-dependent               &$2.864$   &$3.74\times 10^{-3}$  \\
~complex-scaling in Appendix \ref{sec:complexscale} &$2.863$   &$3.82\times 10^{-3}$  \\
~stabilization in Eq. (\ref{eq:CUTEFS})       &$2.863$   &$3.62\times 10^{-3}$  \\
~experiment \cite{NNDC_Chart,2021Wang_AME} &$2.9(3)$  &$-$ \\
\hline  \hline
\end{tabular*}
\end{table}

In Fig. \ref{fig:3702}, the survival probability is displayed.
The fitted result is also presented, where the $1p$-decaying width is obtained as $\Gamma_p=3.74$ keV by assuming the exponential-decaying rule.
This value corresponds to the lifetime of $\tau=\hbar/\Gamma_p \cong 1.76 \times 10^{-19}$ s.
Since the decaying width is small, the survival probability is well approximated as the linear function.
Notice that this width is consistent to the previous result by the stabilization technique.
The present TD calculation inevitably becomes unphysical in the long-time region, $ct \ge 7000$ fm, where
the contamination by reflected waves occurs in the finite box.
For comparison with the present TD-Dirac calculation, I have performed the complex-scaling
calculation \cite{1968Berggren,2014Myo_rev,2022Myo_rev}, which is based on the same Dirac equation.
Note that the same SP potentials and physical parameters are utilized there.
As the result, the $1p$-emission energy and width are obtained as
$Q_p = 2.86$ MeV and $\Gamma_p =3.82$ keV in the $f_{7/2}$ channel.
Details on this complex-scaled Dirac-spinor calculation is separately summarized in Appendix \ref{sec:complexscale}.

Table \ref{table:5324} displays the three sets of results for the $^{36}$Ca$+p$ case.
They are well consistent to each other as obtained by using the same SP potentials.

\modific{3b}{In Appendix \ref{sec:initialtest},
 the sensitivity of TD-Dirac results to the initial setting of confining is examined.
There, I confirmed that the TD-Dirac solution is stable 
as long as the confining barrier is higher than the $1p$-resonance energy.
}

\begin{figure}[b] \begin{center}
  \includegraphics[width = 0.99\hsize]{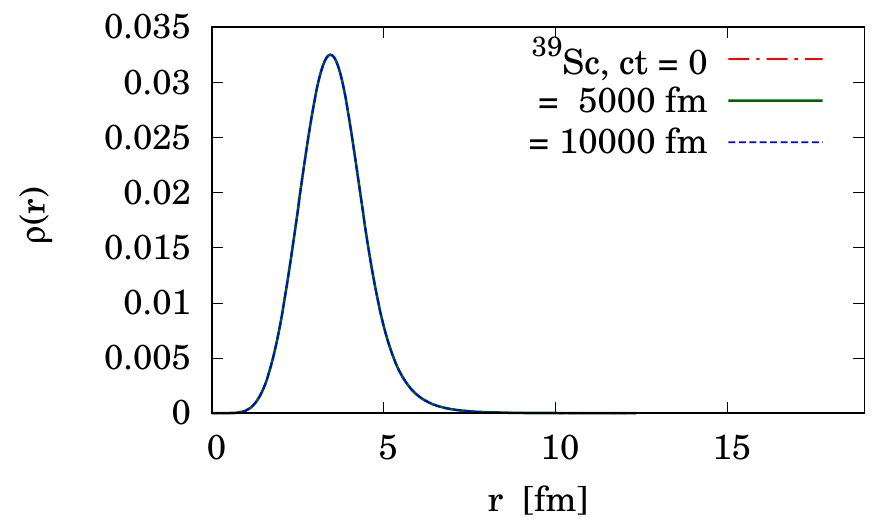}
  \includegraphics[width = 0.99\hsize]{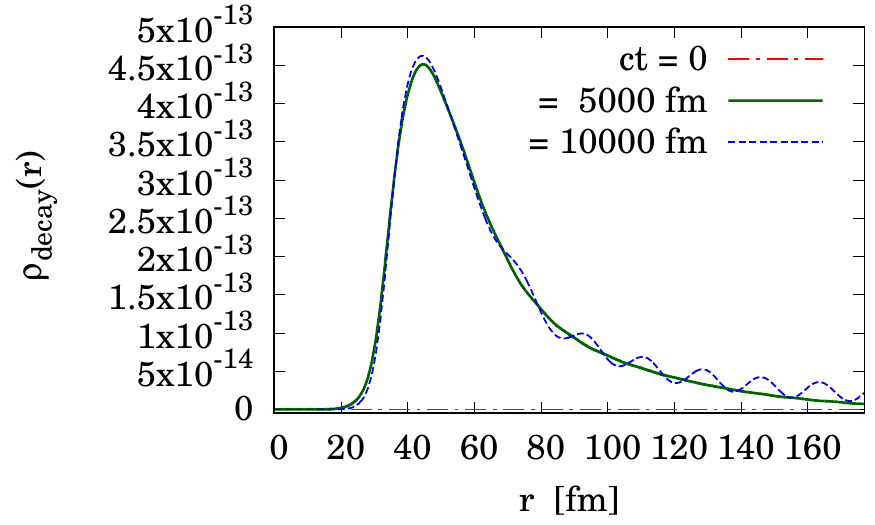}
  \caption{Same to Fig. \ref{fig:0276} but in the $^{39}$Sc case.} \label{fig:0277}
\end{center} \end{figure}

\begin{table*}[t]
\caption{The Q value, $Q_{p}=-S_{p}$, and decaying width of $1p$ emissions evaluated from the TD-Dirac calculations in this work.
The unit is MeV except for the lifetime $\tau =\hbar/\Gamma_{p}$.
\modific{1b}{The results of shell model \cite{96Ormand}, data analysis \cite{96Cole}, as well as experimental data \cite{NNDC_Chart,2021Wang_AME} are presented for comparison.}
}\label{table:2228}
\begin{tabular*}{\hsize} { @{\extracolsep{\fill}} llcclc}
\hline   \hline
&\multicolumn{2}{c}{$^{37}$Sc$= ^{36}$Ca$+p$~~~~~~~~~~}    &&\multicolumn{2}{c}{$^{39}$Sc$= ^{38}$Ca$+p$~~~~~~~~~~}  \\
&~~$Q_{p}$  &$\Gamma_{p}$            &&~~$Q_{p}$  &$\Gamma_{p}$   \\
\hline
~this work (TD-Dirac)  &$2.864$  &$3.74\times 10^{-3}$    &&$0.662$  &$3.87 \times 10^{-9}$   \\
& &($\tau = 1.76\times 10^{-19}$ s)                  &&         &($\tau = 1.70\times 10^{-13}$ s)~ \\
~shell model \cite{96Ormand}    &$2.870(112)$  &$-$   &&$0.639(63)$   &$-$  \\
~data analysis \cite{96Cole}    &$3.006(26)$   &$-$   &&$0.712(26)$   &$-$  \\ 
~experiment \cite{NNDC_Chart,2021Wang_AME}   &$2.9(3)$  &$-$  &~~~~&$0.597(24)$   &($\tau < 400$ ns)  \\
\hline  \hline
\end{tabular*}
\end{table*}

\subsection{Sensitivity to mass numbers}
Next I focus on the other sample case, that is the $^{39}$Sc$= ^{38}$Ca$+p$ system.
Namely, the core nucleus is enlarged from the previous $^{37}$Sc case.
The experimental data give the $1p$-emission Q value, $-S_{p}=597(24)$ keV for this $^{39}$Sc nucleus with the upper limit of lifetime, $\tau < 400$ ns \cite{NNDC_Chart}.
The SP potentials $S(r)$ and $W(r)$ are prepared in the same manner to the previous case.
Namely, the RHB calculation with the same DD-PCX parameters and no-sea approximation
is performed but for this $^{38}$Ca$+p$ case \cite{2014Niksic,2019Yuksel}.
By fitting $S(r)$ and $W(r)$ to this RHB result, I obtained the parameters given in Table \ref{table:SWPC}.
I have confirmed that SP levels up to the $0d_{3/2}$ are bound for the $Z=20$ protons.
The time-development calculation with the confining potential is then repeated.
The initial state is solved to have the mean Q value of $-S_{p}=662$ keV,
which is slightly higher than the experimental value.

In Fig. \ref{fig:3702}, the survival probability of the $^{39}$Sc nucleus is presented.
The $1p$-decaying width is evaluated as $\Gamma_{p} =3.87 \times 10^{-9}$ MeV,
which is remarkably reduced from the previous $^{37}$Sc $\cong ^{36}$Ca$+p$ case.
Note that the corresponding lifetime is given as $\tau \cong 1.7 \times 10^{-13}$ s, which is consistent to the experimental limitation.
\modific{1c}{In Table \ref{table:2228}, my TD-Dirac results for $^{37}$Sc and $^{39}$Sc are summarized.
One can find the sensitivity of $1p$ emission to mass numbers.
Notice also that the present results are in good agreement with the non-relativistic shell-model calculations \cite{96Ormand}.
}

\begin{figure}[b] \begin{center}
  \includegraphics[width = 0.99\hsize]{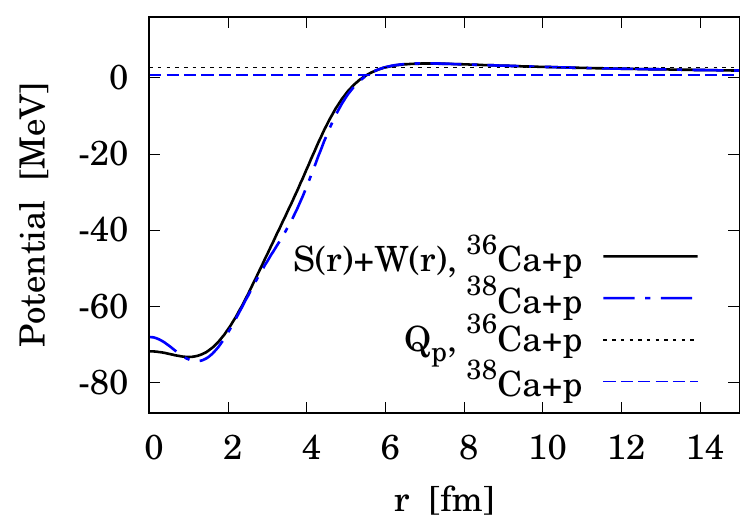}
  \caption{The SP potentials used to simulate the time-dependent $1p$ emissions from $^{37}$Sc and $^{39}$Sc nuclei.
Those are fitted to the RHB results with the DD-PCX parameters \cite{2014Niksic,2019Yuksel}.
The $1p$-emission Q values obtained with time-dependent calculations are also plotted by dotted lines.} \label{fig:6475}
\end{center} \end{figure}

\modific{}{The sensitivity to mass numbers
or equivalently to the size of system can be understood from the profile of SP potentials.}
In Fig. \ref{fig:6475}, the total potentials $S(r)+W(r)$ for the $^{37}$Sc and $^{39}$Sc nuclei are compared.
One can read that, in the $^{39}$Sc case, its potential has the deeper profile around $r\cong 4$ fm.
This is naturally understood from the larger size of the core nucleus.
In correspondence, the SP resonance becomes deeper with smaller $Q_p$ than the $^{37}$Sc case.
On the other side, the Coulomb barrier around $r \cong 6$ fm shows the similar form
consistently to the common atomic number of the two nuclei.
Since the quantum-tunneling effect is enhanced (reduced) with the higher (lower) SP energy against the same potential barrier,
the $1p$ emission of the $^{39}$Ca case has the longer lifetime.
This conclusion is qualitatively consistent to Refs. \cite{2016TTSun,2019Wang_Dirac,2016Kobayashi}.

\modific{6}{For quantum-mechanical decaying processes, in general, 
the deviation from the exponential decay occurs in the extremely long-time region \cite{1961Winter,1977Nico,2002Dicus,2021Jimenez}.
This long-time development, however, is beyond the present scope,
due to the contamination of reflected waves.
In addition, I also confirmed that, in Fig. \ref{fig:3702}, the survival probability shows the other kind of deviation from the exponential decay in the early stage.
That is, by symbolically writing,
$P_{\rm surv}(t) \cong  1- \alpha \left( \frac{t}{\tau} \right)^2$ for $t \ll \tau$.
This polynomial behaviour has been well known, and discussed with interests not only of nuclear physics
but also of generally quantum-mechanical time-dependent processes,
being relevant to the quantum Zeno effect \cite{1977Misra,1977Chiu,1988Levitan,2021Jimenez}.
}

Since the decaying width is extremely narrow in the $^{39}$Sc case,
the other two methods, stabilization and complex scaling,
cannot give a clear result in this work.
In the stabilization method including this narrow resonance, the fitting procedure
needs the corresponding accuracy, for which several technical problems remain.
The complex-scaled Dirac-spinor calculation, on the other side,
has not found the complex-eigen energy with a finite width: the calculation
inevitably converges to $\Gamma_p =0$, whereas the real part can be reproduced
as $Q_p=+0.663$ MeV consistently to the TD-Dirac method.
One possible reason is that it needs the fine mesh of complex coordinates to reduce numerical errors \cite{1989Atkinson},
when the expected width is small.
Because of computing cost, this task is left for the future improvement.

\section{Summary} \label{Sec:Sum}
I investigate the quantum-tunneling effect on the $1p$-radioactive emission based on the Dirac equation.
As one tool to evaluate the $1p$-decaying width, which physically corresponds to the tunneling probability,
the time-dependent calculation of Dirac spinor is utilized.
This method is applied to $1p$ emissions from the $^{37}$Sc and $^{39}$Sc nuclei,
which can be well approximated as the valence proton and the core nuclei.
By utilizing the RHB calculation, the SP potentials for the TD-Dirac spinor are determined.
The sensitivity of $1p$-emission energy and decaying width to the mass number is demonstrated \cite{2016TTSun, 2019Wang_Dirac}.
This is because the size of system is reflected on the nuclear part of SP potentials,
whereas the Coulomb barrier is common due to the same atomic number.
The calculated $1p$ energy and decaying lifetime are roughly consistent to the experimental limitation.

In this paper, I only discuss the Ca cores,
\modific{4b}{where the proton-shell closure at $Z=20$ enables me to work with the valence-proton approximation for emission.
For open-shell systems, the pairing correlation requires a more careful treatment \cite{03Dean_rev,2009Oba,2019TTSun,2022Cao}.
Instead of real-particle resonance, the time development of quasi-particle resonance should be taken into account \cite{2001Grasso,2003Hamamoto,2016Kobayashi,2020Kobayashi}.
}
Even though this task remains for future progress,
the present TD method is expected as applicable to the other proton-drip lines.
The evaluation of $1p$ radioactivity with various REDFs for other systems is in progress now.
Checking the consistency of TD calculations with other methods \cite{2016Niu_Dirac,2017Fang_Dirac,2019Wang_Dirac,2022Cao,2010Li_Dirac,2014TTSun,2016TTSun,2020TTSun} is also necessary to avoid the methodological biases.
Note that, for several proton-rich nuclei, the experimental access is still challenging, and thus,
the improvement of predicting ability is on demand.
Considering that TD calculation enables one to intuitively understand the dynamics,
its application to the two-proton radioactivity could be beneficial for the improvement of REDF \cite{2009Gri_rev,2012Pfu_rev,13Olsen,2019Qi_rev}.
For this purpose, however and again, one needs to take the pairing correlation into account \cite{2017Oishi,2021Wang_Naza}.
Also, the relative motion between two emitted protons requires a large increase of computing cost.
This project still waits for several technical developments.

\modific{2}{In this work,
the Runge-Kutta method is employed to solve the wave functions of single-proton resonance.
One may consider more general cases, including meta-stable states with the multi-body structure, multi-channels of resonance, and/or the scattering processes.
In such cases, the Runge-Kutta method could not be appropriate, since the asymptotic behaviour is not trivial, and thus, numerical precision is not obviously guaranteed.
One alternative option may be the basis-expansion method.
There, however, one must carefully choose or newly build up the basis to reproduce, e.g. the spatial distribution in the asymptotic region \cite{2009Gri_rev,01Gri_I,2021Wang_Naza}.
The other possibility is the lattice solutions, including the space-lattice \cite{2012Bulgac,2016Sekizawa,2022Ren_TDGCM} and momentum-lattice ones \cite{2016Niu_Dirac,2017Fang_Dirac,2019Wang_Dirac,2022Cao}.
}

\section*{Acknowledgment}
This work is supported by the Yukawa Research Fellow Programme by Yukawa Memorial Foundation in Kyoto University.
I sincerely thank Nils Paar, Tomoya Naito, Akira Ohnishi, and Takayuki Myo for fruitful discussions.

\appendix

\section{Complex-scaled Dirac-spinor method} \label{sec:complexscale}
For comparison with TD-Dirac calculations in the main sections,
here I introduce the complex-scaling calculations based on the Dirac formalism.
Details on the complex-scaling method are well summarized in Refs. \cite{2014Myo_rev,2022Myo_rev}.
This method has been utilized to describe a variety of nuclear meta-stable states.
The basic idea starts from the arbitrary potential problem,
which is in many cases the Schr\"{o}dinger or Dirac equation:
$\oprt{H} \psi(\bir) = E \psi(\bir)$, where $\oprt{H}$ includes the potential $V(\bir)$.
For the spherical system, the complex scaling simply reads
\beq
U(\theta):~~r \longrightarrow r \exp (i \theta),
\eeq
where $\theta$ is the complex-scaling angle.
In this way, the Hamiltonian as well as the state are also transformed as \cite{2014Myo_rev,2022Myo_rev}
\beq
\oprt{H}^{\theta} = U(\theta) \oprt{H} U^{-1}(\theta),~~~\psi^{\theta} = U(\theta)\psi.
\eeq
For solving the typical resonance of $\epsilon = E -i\Gamma/2$,
one needs to set $\theta > \frac{1}{2} \arctan \left( \frac{\Gamma}{2E} \right)$
in numerical calculations \cite{1997Myo}.

\begin{figure}[tb] \begin{center}
  \includegraphics[width = 0.99\hsize]{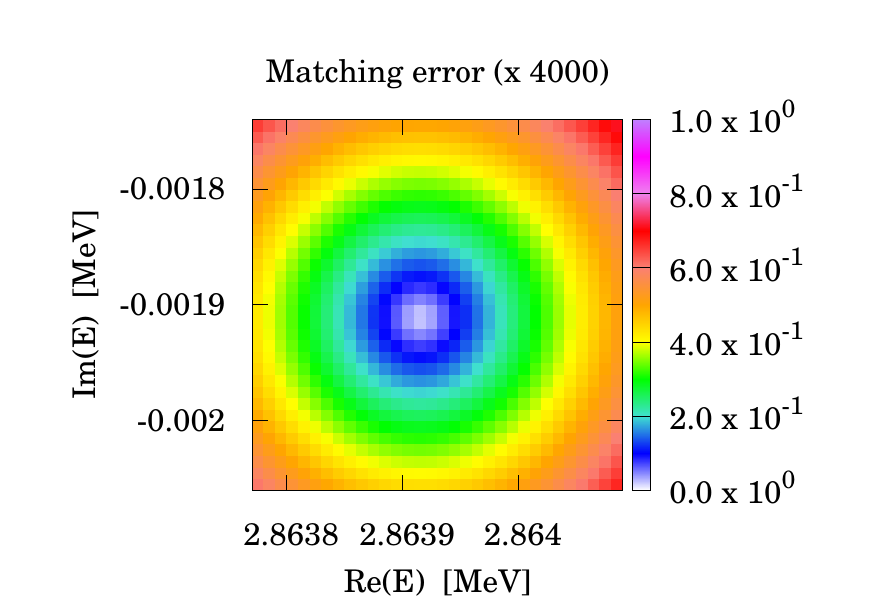}
  \caption{The error of matching, $X(\epsilon_N,r_m,r_n)$, where $r_m=4.1$ fm and $r_n=8.2$ fm. Plotted in the arbitrary scale.} \label{fig:CA652}
\end{center} \end{figure}

At the level of numerical calculations in this work,
I utilize the complex-scaled Runge-Kutta (CSRK) method.
Namely, the Dirac equation given in Eq. (\ref{eq:CF29KQ}) but after the complex scaling is computed with the Runge-Kutta method:
\beq
\frac{d}{dz}
\left( \begin{array}{c} a_N(z) \\ b_N(z) \end{array}  \right)
= \left( \begin{array}{cc} h_{11} & h_{12}  \\  h_{21} & h_{22} \end{array} \right)
\left( \begin{array}{c} a_N(z) \\ b_N(z) \end{array} \right),
\eeq
where $z=r\exp(i\theta)$,
$h_{11}=\frac{\kappa_{lj}}{z}$,
$h_{12}=\frac{s(z)+v(z,\epsilon_N)}{\hbar c}$,
$h_{21}=\frac{s(z)-v(z,\epsilon_N)}{\hbar c}$, and
$h_{22}=-\frac{\kappa_{lj}}{z}$.
Note that the real (imaginary) part of the eigen energy, $\epsilon_N=Q_p-i\Gamma_p /2$,
is interpreted as the Q value (width) of the $1p$-emission in the present case.
Numerical calculations are performed with the same SP potentials used
for $^{37}$Sc$= ^{36}$Ca$+p$ in the main text.
I focus on the $f_{7/2}$ channel.

The complex-scaling method enables one to solve the bound and resonant states in the common manner.
Thus, for finding the complex-eigen energy $\epsilon_N$,
one can use the same technique of matching, namely, the wave function and its derivative
need to match between the forward and backward solutions.
In this paper, at the matching point $r_m$, the {\it error} of matching is determined as
$W(\epsilon_N, r_m) = a'_{F} a_{B} - a_{F} a'_{B}$,
where $a_F$ and $a'_F$ ($a_B$ and $a'_B$) are the forward (backward) solution and its derivative, respectively, for the larger component, $a_{N}(z_m=r_m e^{i\theta})$.
As one technique, I refer to the two points for matching, $r_m=4.1$ fm and $r_n=8.2$ fm.
That is
\beq
X(\epsilon_N,r_m,r_n) = \sqrt{\abs{W(\epsilon_N,r_m)} \cdot \abs{W(\epsilon_N,r_n)}}. \label{eq:FS63IF}
\eeq
The complex-eigen energy, $\epsilon_N=Q_p-i\Gamma_p /2$, is solved so as to minimize this quantity.

\begin{figure}[tb] \begin{center}
\includegraphics[width = 0.99\hsize]{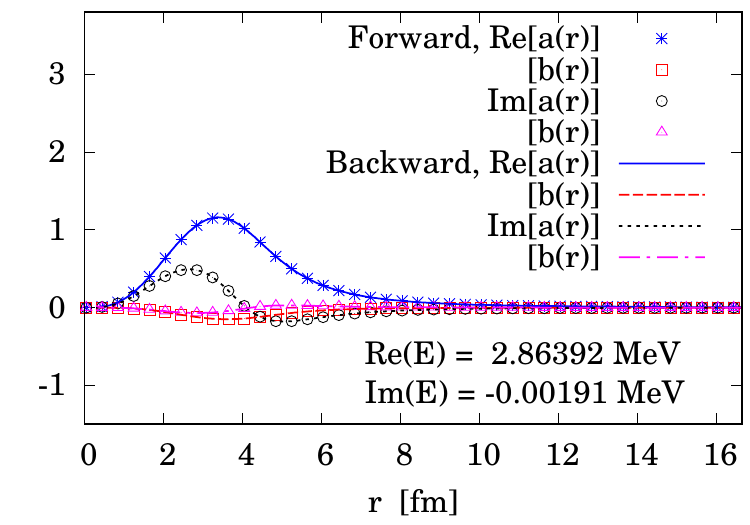}
\includegraphics[width = 0.99\hsize]{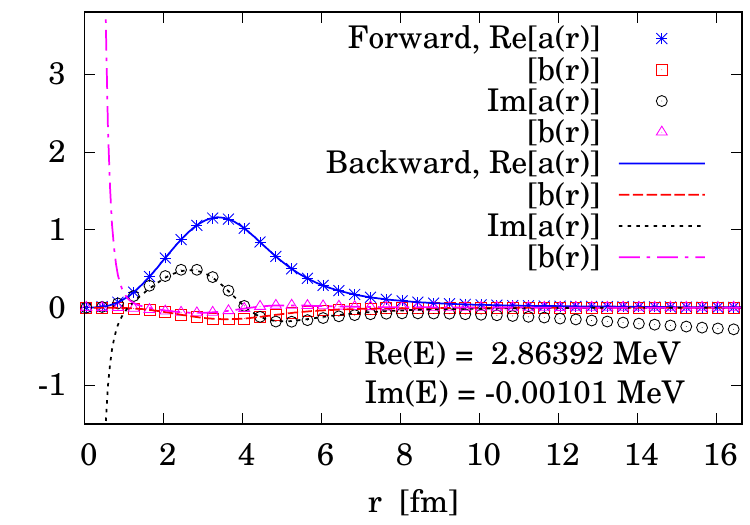}
\caption{(Top) The complex Dirac spinor $a_N(z)$ and $b_N(z)$ obtained with CSRK method for $Q_p=2.86392$ MeV and $\Gamma_p/2 = -0.00191$ MeV.
System is $^{37}$Sc$= ^{36}$Ca$+p$.
The complex coordinate is determined as $z=r \exp(i \theta)$ in numerical calculations.
Plotted in the arbitrary scale.
(Bottom) The same plot but for $\Gamma_p/2 = -0.00101$ MeV.} \label{fig:CA653}
\end{center} \end{figure}

\begin{figure}[tb] \begin{center}
\includegraphics[width = 0.99\hsize]{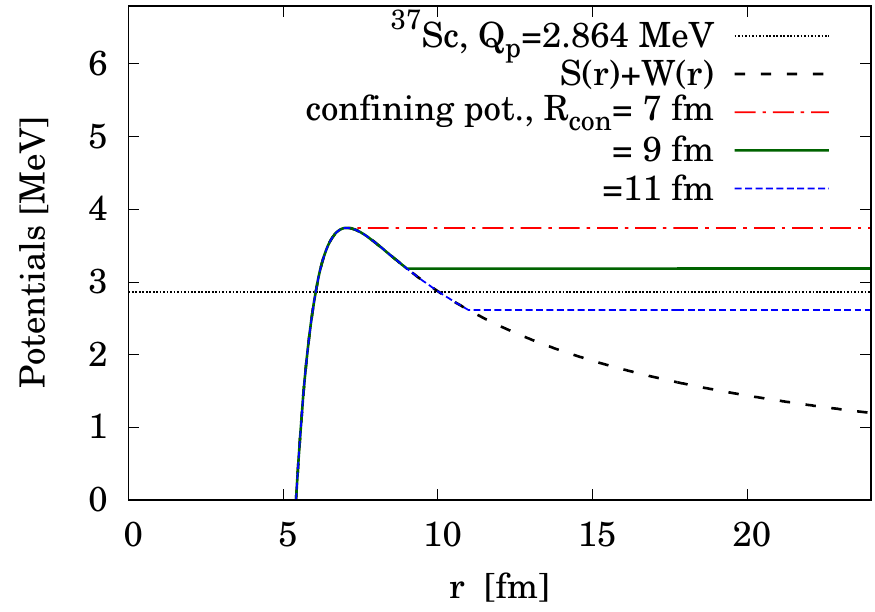}
\includegraphics[width = 0.99\hsize]{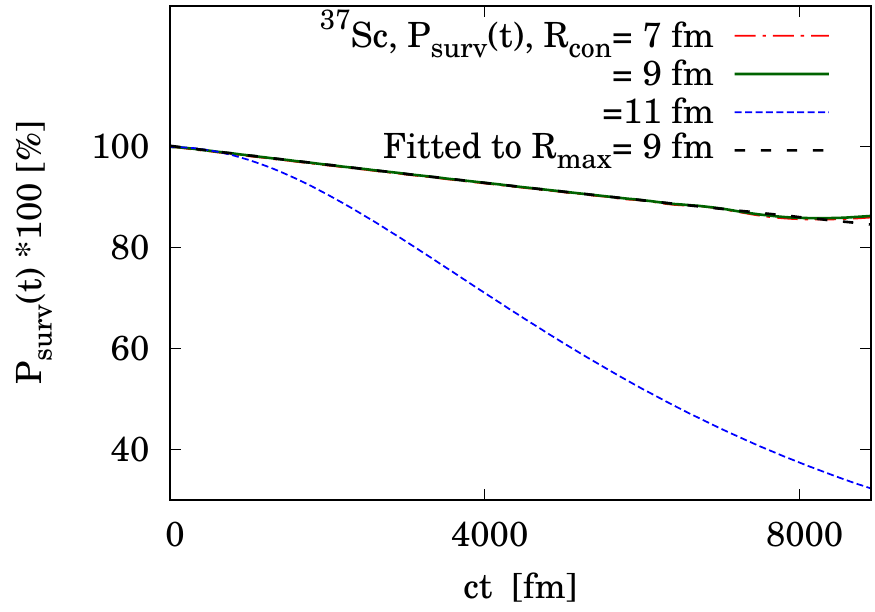}
\caption{(Top) Confining potentials used to check the initial-state dependence in the $^{36}$Ca+$p$ case.
(Bottom) The survival probabilities obtained with various confining potentials.} \label{fig:VOU6S11}
\end{center} \end{figure}

\begin{figure}[tb] \begin{center}
\includegraphics[width = 0.99\hsize]{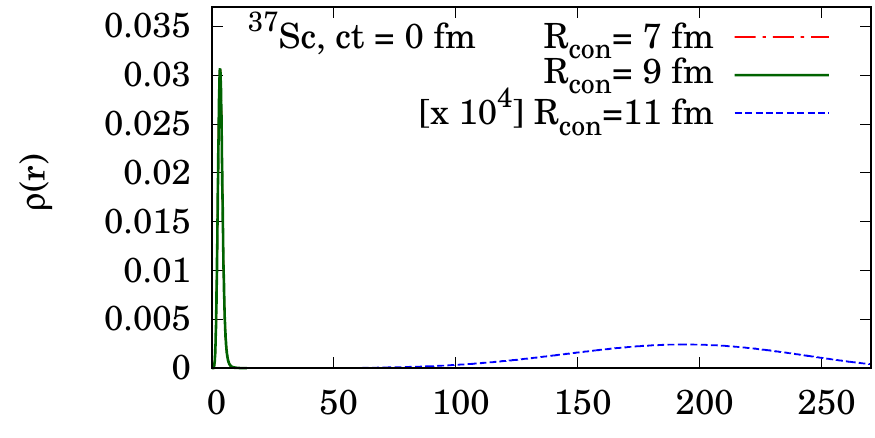}
\includegraphics[width = 0.99\hsize]{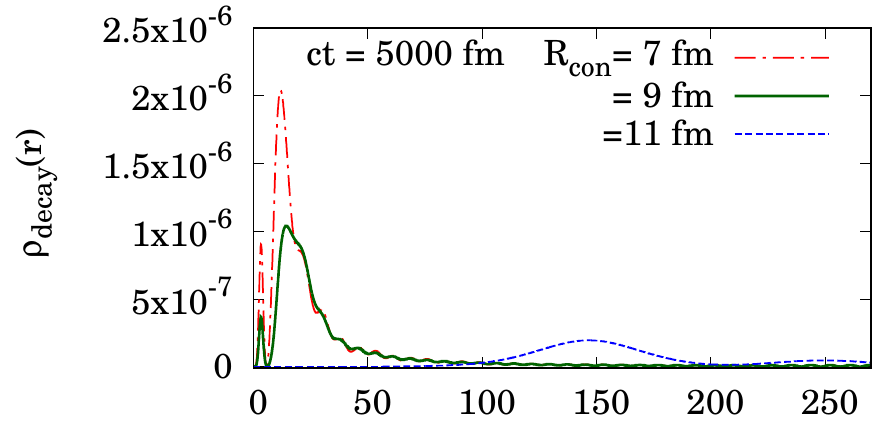}
\caption{(Top) The one-proton densities at $t=0$ obtained with the confining potentials in Fig. \ref{fig:VOU6S11}.
(Bottom) Same plotting but for the decaying states at $ct=5000$ fm.} \label{fig:VOU6S12}
\end{center} \end{figure}

In Fig. \ref{fig:CA652}, the error of matching is plotted as a function of real and imaginary parts of the complex-eigen energy.
The minimum is found at $Q_p=2.86392$ MeV and $\Gamma_p/2 = -0.00191$ MeV.
The corresponding solutions of Dirac spinor, $a_N(z)$ and $b_N(z)$, are presented in Fig. \ref{fig:CA653},
where both the real and imaginary parts well agree between the forward and backward solutions.
For comparison, in Fig. \ref{fig:CA653}, I also plot the same results but slightly changing the width input.
There, at $r\cong 0$ and $r >8$ fm, several components become diverged between the forward and backward solutions.

\modific{5}{As described in the main text, for the $^{36}$Ca+$p$ case, the TD-Dirac and complex-scaling calculations are in good agreement.
Note that, if the lifetime was extremely short, the time-dependent simulation could have a risk of divergence,
whereas the complex-scaling method can generally apply to such a broad-width case \cite{2014Myo_rev,2022Myo_rev}.
When one changes the target from the single-body to the multi-body meta-stable systems,
the computing cost of time development is expected as higher than the complex-scaling method.
In parallel, I notify that the complex-scaling method in principle solves the single
set of real and imaginary values of the energy: $\oprt{H} \ket{\psi} = \left( E-i\Gamma/2 \right) \ket{\psi}$.
Thus, the decaying rule of this eigenstate is purely exponential, where deviations in the short and long-time regions do not appear.
For several observable quantities, their expectation values evaluated within the complex-scaling method inevitably become complex, where the physical interpretation is not obvious \cite{2014Myo_rev,2022Myo_rev}.
Although these problems remain for future discussions,
the present comparison of TD-Dirac and complex-scaling methods supports}
their consistency.

\section{Dependence on the initial condition} \label{sec:initialtest}
\modific{3}{In the main sections,
for the $^{36}$Ca+$p$ case, I utilized the confining potential for $r \ge R_{\rm con}=9$ fm.
In time-dependent simulations, there is often a risk that the choice of initial state may provide the unphysical results.
Therefore, in this section, the dependence of TD-Dirac results on the initial condition is examined.
For this purpose,
I perform the same calculations but by changing the confining radius as $R_{\rm con}=7$ and $11$ fm.
See also Refs. \cite{94Serot,00Talou} for the similar studies based on the Schr\"{o}dinger formalism.}

Figure \ref{fig:VOU6S11} displays the three cases of $R_{\rm con}=7$, $9$ (default), and $11$ fm.
The Coulomb barrier exists at $r\cong 6$ fm in the original potential $S(r)+W(r)$.
Note that the $1p$-emission energy was obtained as $Q_{p}=2.864$ MeV in the default case.
First, by checking the survival probability $P_{\rm surv}(t)$,
the result keeps unchanged as long as the wall potential is above the $1p$-emission energy 
with $R_{\rm con}=7$ and $9$ fm.
With these two well-confining potentials, the decaying width
is commonly evaluated as $\Gamma_{p}=3.74$ keV with the fitting procedure for $P_{\rm surv}(t) \propto e^{-t\Gamma_p /\hbar}$.
Then, when this wall becomes lower than $Q_{p}$ with $R_{\rm con}=11$ fm, the survival probability drastically changes.
The decaying lifetime with this shallow confining looks much shorter than the former two cases.
This result is, however, not of the proper $1p$-radioactive process anymore.
The $1p$ energy, $Q_p=\Braket{\psi(0) \mid \oprt{H}_{D} \mid \psi(0)}$, is obtained as $2.865$, $2.864$, and $0.156$ MeV
with $R_{\rm con}=7$, $9$, and $11$ fm, respectively.
Therefore, the shallow-confining potential fails to reproduce the $1p$ resonance.

In Fig. \ref{fig:VOU6S12}, the $1p$ densities of three initial states are plotted.
The shallow-confining case with $R_{\rm con}=11$ fm shows a wide distribution,
meaning that the valence proton is already outside.
Thus, this initial state is not suitable to the radioactive emission.
For the two other cases, the initial density is well localized inside the barrier.
In the bottom panel of Fig. \ref{fig:VOU6S12},
the densities of decaying states, $\ket{\psi_{\rm decay}(t)}$ in Eq. (\ref{eq:psidecay}),
are also plotted at $ct=5000$ fm.
The two well-confining cases show a similar pattern of decaying density consistently to the common value of decaying width.
Consequently, as long as the confining barrier is higher than the expected energy of resonance,
the present TD-Dirac calculation yields the stable results.


%

\end{document}